\documentclass{aa}  
\usepackage{graphicx}
\usepackage{txfonts}
\usepackage{epsfig}
\begin{document}
   \title{The optical counterpart of IGR J00291+5934 in quiescence\thanks{Based on observations made with the Italian Telescopio Nazionale Galileo (TNG) 
   operated on the island of La Palma by the Fundaci\'on Galileo Galilei of the INAF (Istituto Nazionale di Astrofisica) at the Spanish Observatorio del Roque 
   de los Muchachos of the Instituto de Astrofisica de Canarias.}}


   \author{P. D'Avanzo
          \inst{1, 2}
	  \and
          S. Campana\inst{1}
	  \and
          S. Covino\inst{1}
          \and
          G. L. Israel\inst{3}
          \and
          L. Stella\inst{3}
          \and
          G. Andreuzzi\inst{4}
          }

   \offprints{P. D'Avanzo}

   \institute{INAF, Osservatorio Astronomico di Brera, via E. Bianchi 46, I-23807 Merate (Lc), Italy\\
              \email{paolo.davanzo@brera.inaf.it}
         \and
             Universit\`a degli Studi dell'Insubria, Dipartimento di Fisica e Matematica, via Valleggio 11, I-22100 Como, Italy
         \and
             INAF, Osservatorio Astronomico di Roma, via Frascati 33, I-00040 Monte Porzio Catone, Roma, Italy
         \and
             Fundaci\'on Galileo Galilei - INAF, Fundaci\'on Canaria, C. Alvarez de Abreu, 70, 38700 S/C. de La Palma, Spain
             }

   \date{Received; accepted}

 
  \abstract
   {}
   {The recent (December 2004) discovery of the sixth accretion-powered
millisecond X-ray pulsar IGR J00291+5934 provides a very good chance to 
deepen our knowledge of such systems. Although these systems are well studied at high energies, 
poor informations are available for their optical/NIR counterparts during quiescence.
Up to now, only for SAX J1808.4-3658, the first discovered system of this type, we have a 
secure multiband detection of its optical counterpart in quiescence. Among the seven known system 
IGR J00291+5934 is the one that resembles SAX J1808.4-3658 more closely.}
   {With the Italian 3.6 m TNG telescope, we have performed deep optical and NIR photometry of the field of 
   IGR J00291+5934 during quiescence in order to look for the presence of a variable counterpart.}
   {We present here the first multiband ($VRIJH$) detection of the optical and NIR counterpart of 
IGR J00291+5934 in quiescence as well as a deep upper limit in the $K-$band. We obtain an optical   
light curve that shows variability consistent with a sinusoidal modulation at the known 2.46 hr orbital 
period and present evidence for a strongly irradiated companion.}
   {}


   \maketitle
%

\section{Introduction}

In 1998, the discovery of the first accretion$-$powered millisecond X$-$ray pulsar SAX J1808.4$-$3658 (Wijnands \& van der Klis 1998; Chakrabarthy \& Morgan
1998), confirmed the evolutionary link between Low Mass X$-$Ray Binaries (LMXBs) and millisecond radio 
pulsars (see e.g. Bhattachrya \& van den Heuvel 1991; Tauris \& van den Heuvel 2004) supporting the idea that the formers are the progenitors of the latters.
In the following years, six more accretion$-$powered millisecond X$-$ray pulsars have been 
discovered: XTE J1751$-$305 (Markwardt et al. 2002), XTE J0929$-$314 (Galloway et al. 2002), XTE J1807$-$294 (Markwardt, Smith \& Swank 2003; Campana et al.
2003), XTE J1814$-$314 (Markwardt \& Swank 2003; Strohmayer et al. 2003), IGR J00291+5934 (Galloway et al. 2005) and HETE J1900.1-2455 
(Morgan et al. 2005; Campana 2005). 
All these systems are transients of the Soft X$-$Ray Transients class (SXRTs, for a review see~\cite{Ca98}), have orbital periods in the range 
between 40 min and 4.5 hr and spin frequencies from 1.7 to 5.4 ms. These seven accreting millisecond pulsars are well studied at high 
energies, especially in the X$-$rays, both in outburst and in quiescence (see \cite{Wi05b} for a review). 
On the other hand, with the significant exception of SAX J1808.4$-$3658, 
their optical/NIR quiescent counterparts are only poorly known. The optical light curve of 
SAX J1808.4$-$3658 in outburst and quiescence shows variability modulated at
the orbital period, in antiphase with the X$-$ray light curve (\cite{Ho01}; \cite{Ca04}). 
This is unlike other quiescent transient that normally show a double-humped morphology, due to an 
ellipsoidal modulation, and indicates that the companion star is subject to some irradiation. Burderi et al. (2003) proposed that the irradiation is due to 
the release of rotational energy by the fast spinning neutron star, switched on, as a radio pulsar, during quiescence. Following this idea, Campana et al. (2004) 
measured the required irradiating luminosity needed to match the optical flux and found that it is a factor of about 100 larger than the quiescent X$-$ray luminosity of the
system. Neither accretion$-$driven X$-$rays nor the intrinsic luminosity of the secondary star or the disc can account for it. So, these authors conclude that the
only source of energy available within the system is the rotational energy of the neutron star, reactivated as a millisecond radio pulsar.
Optical and NIR observations performed in the past by different groups for the other systems of this class only led to deep upper limits for the counterparts of 
XTE J1751$-$305 (\cite{Jo03}) and XTE J1814$-$314 (\cite{Kr05}) or to the detection of very faint candidates, if any (\cite{Mon05} 
for XTE J0929$-$314). The intrinsic faintness of the targets, in combination with the high interstellar absorption and high stellar crowding of the
relevant fields are among the main reasons for the lack of detections at optical wavelengths.


\section{IGR J00291+5934}

IGR J00291+5934 was discovered in outburst
on December 2nd 2004 during a routine monitoring of the Galactic plane with the
\textit{INTEGRAL} satellite (\cite{Ec04}). Follow-up \textit{RXTE} observations established the source as a 
598.88 Hz (1.67 ms) accretion-powered X--ray pulsar (\cite{Ma04a}) with 
a 2.46 hr orbital period (\cite{Ma04b}). With a spin frequency of 599
Hz, IGR J00291+5934 is the fastest known accretion-powered millisecond pulsar.
An $R \sim$ 17.4 candidate optical counterpart was identified within the 
\textit{INTEGRAL} error circle (\cite{Fo04}). Optical spectroscopy 
of the candidate revealed weak He II and H$\alpha$ emission, supporting its association 
with IGR J00291+5934 (\cite{Ro04}). The mass function gives an upper limit of 0.16 $M_{\odot}$ for the mass of the companion
star indicating that, like SAX J1808.4$-$3658, the secondary is probably a hot brown dwarf (\cite{Ga05}). Radio observations 
provided evidence for variable emission, consistent with the counterpart position (\cite{Po04}; \cite{Fe04}).
Signs of source activity during two occasions in the past were found in the $RXTE$/ASM light curves, with a probable
recurrence of about 3 years (\cite{Re04}). Observations made with $INTEGRAL$ and $RXTE$ during outburst revealed a power-law spectrum (with photon index $1.7
- 1.8$) consistent with a column density of $N_H$ $\sim 10^{22}$ cm$^{-2}$ (\cite{Sh05}; \cite{Ga05}). Paizis et al. (2005) performed
$Chandra$ and $RXTE$ spectroscopy during outburst and modeled the spectra with a combination of a thermal component (probably due to a hot-spot on the
neutron star's surface) and a power-law with a hydrogen column of $N_H = 4.3 \times 10^{21}$ cm$^{-2}$. 
In quiescence IGR J00291+5934 has been recently detected by $Chandra$ 
(and historically ROSAT; \cite{Jo05}). The source quiescent 0.5--10 keV unabsorbed flux was 
$\sim 8 \times 10^{-14}$ erg cm$^{-2}$ s$^{-1}$ (with $N_H$ fixed to $2.8 \times 10^{21}$ cm$^{-2}$) during two $Chandra$ 
observations following the 2004 outburst. However, in the third observation a few weeks later, IGR J00291+5934 showed a flux a factor of two higher. 
This yielded the first clear evidence for variability in the quiescent X$-$ray emission of an accretion$-$powered millisecond X$-$ray pulsar.
Subsequent quiescent $Chandra$ observations carried out in November 2005 showed a power -law spectrum (with photon index $2.0 - 2.9$) and unabsorbed $0.5-10$
keV flux of $\sim 7 \times 10^{-14}$ erg cm$^{-2}$ s$^{-1}$ for $N_H$ $= 4.6 \times 10^{21}$ cm$^{-2}$. Very recently, Torres et al. (2007) reported about
optical, NIR and $Chandra$ observations of IGR J00291+5934 carried out during outburst as well as in quiescence. They model the quiescent $0.5-10$ keV spectrum with a
power$-$law with a photon index of $\sim 2.4$ and $N_H = (4 - 5) \times 10^{21}$ cm$^{-2}$. 

The non$-$detection of X$-$ray bursts (\cite{Ga05}) makes the estimate of the distance of IGR J00291+5934 difficult.
Shaw et al. (2005) give an upper limit of 3.3 kpc considering source position with respect to the Galactic Centre and the
measured absorbing column density\footnote{Those authors used $N(H) = (2.8 \pm 0.4) \times 10^{21}$ cm$^{-2}$ as measured by Nowak et al. (2004) 
with $Chandra$. Even if this value has been largely used in the literature, we note that from a subsequent analysis of the same $Chandra$ 
data, the same group measured a value of $N(H) = (4.3 \pm 0.4) \times 10^{21}$ cm$^{-2}$ (as reported in \cite{Pa05}). Using this refined value of
$N(H)$ and with the same arguments of Shaw et al. (2005) we derive for the distance of IGR J00291+5934 an upper limit of $\sim 6.0$ kpc.}. 
From the $INTEGRAL$ observed total fluence and taking into account the recurrence time of 3
years, Galloway et al. (2005) and Falanga et al. (2005) find a minimum distance of, respectively, 4 and 4.7 kpc. By 
considering the non$-$detection of thermonuclear burst, Galloway et al. (2005) suggest that the source distance should not be 
significantly greater than 4 kpc. Jonker et al. (2005) estimate the source distance to be between 2.6--3.6 kpc 
assuming a quiescent X$-$ray luminosity of $(5-10) \times 10^{31}$ erg s$^{-1}$, similar to that of SAX J1808.4$-$3658 and XTE
J0929$-$314 (\cite{Ca02}; \cite{Wi05a}). Torres et al. (2007) estimated a distance of $2-4$ kpc based on an estimate of the critical 
X$-$ray luminosity necessary to ionize the accretion disc and produce the observed X$-$ray light curve during outburst. Finally, Burderi et al. 
(2006) derive the mass accretion rate from timing of the source and estimate a bolometric luminosity of about $10^{38}$ erg s$^{-1}$ 
concluding that the source is in a distance range of $7-10$ kpc. They also note that 10 kpc is close to the edge of our Galaxy in the 
direction of IGR J00291+5934. In the following, we assume a fiducial distance of 3 kpc.


\section{Observations and data reduction}

Optical and NIR observations of the field of IGR J00291+5934 were carried out with the Italian 3.6-m 
TNG telescope, sited in Canary Island, in 2005 on three different nights. Optical observations were made using
the DOLORES (Device Optimized for the LOw RESolution) camera, installed at the Nasmyth$-$B focus 
of the TNG. The detector is a Loral thinned and back-illuminated 2048x2048 CCD. The scale is 0.275 arcsec/pix which yields 
a field of view of about 9.4 x 9.4 arcmin. A set of $V$ and $R-$band images was taken on 
2005 August 11 and a set of $I-$band images was taken on November 3. The exposure time of each frame was
260 s, with a detector dead time of about 70 s. With each optical filter we covered about $80-90$\% of the 2.46 hr orbital
period, with approximately a 0.04 phase resolution. NIR observations were performed on 2005 September 2nd using the  NICS 
(Near Infrared Camera Spectrometer) infrared camera, which is based on a HgCdTe Hawaii 1024x1024 array ($4.2 \times 4.2$ arcmin). 
The instrument is installed at the Nasmyth$-$A focus of the TNG. We cycled through $JHK$ filters with exposures times of 20
and 30 s, obtaining a total exposure of about 30 min for each filter. The complete observing log is presented in Tab.~\ref{tab:log} 

Image reduction was carried out by following the standard procedures: subtraction of an averaged bias frame,
division by a normalized flat frame. Astrometry was performed using the USNOB1.0\footnote{http://www.nofs.navy.mil/data/fchpix/}
and the 2MASS\footnote{http://www.ipac.caltech.edu/2mass/} catalogues.
PSF-photometry was made with the ESO-MIDAS\footnote{http://www.eso.org/projects/esomidas/} daophot task for all the objects in the 
field. The calibration was done against Landolt standard stars for $V$ and $R$ filters and against the 2MASS 
catalog for NIR filters. $I-$band observations were carried out under non-photometric conditions, so we could not calibrate them using Landolt standard 
stars as reference. We thus used the USNOB1.0 catalog to calibrate the $I$ frames exploiting, to this end, a large sample of local isolated and non-saturated 
field stars. This led to a slight increase in the uncertainty of the magnitude measurement, even if $I-$band images were taken under good seeing
conditions (see Tab.~\ref{tab:log}).

In order to minimize any systematic effect, we performed differential photometry with respect to a
selection of local isolated and non-saturated standard stars. In addition to ordinary photometry, we also carried out image subtraction with the
ISIS package (\cite{Al00}; \cite{Al98}) in order to check for variability of the sources in the field.

\begin{table}
\caption{Observation log for IGR J00291+5934.}
\begin{tabular}{ccccc}
\hline
UT observation &  Exposure &   Seeing     &   Instrument       & Filter\\
  (YYYmmdd)    &  (s)      &  (arcsec)    &                    &       \\ 
\hline
20050811.06237 &  $23 \times 260$ s            & $1.4''$  & TNG/LSR       & $V$	 \\
20050811.17013 &  $24 \times 260$ s            & $1.2''$  & TNG/LSR       & $R$	 \\
20051103.07361 &  $23 \times 260$ s            & $0.9''$  & TNG/LSR       & $I$	 \\
20050902.04549 &  $30 \times 3 \times 20$ s    & $0.7''$  & TNG/NICS      & $J$	 \\
20050902.06426 &  $20 \times 3 \times 20$ s    & $0.7''$  & TNG/NICS      & $H$	 \\
20050902.06983 &  $20 \times 3 \times 20$ s    & $0.8''$  & TNG/NICS      & $K$	 \\
\hline
\end{tabular}
\label{tab:log}
\end{table}

%

\section{Results}

\begin{figure}
\epsfig{file=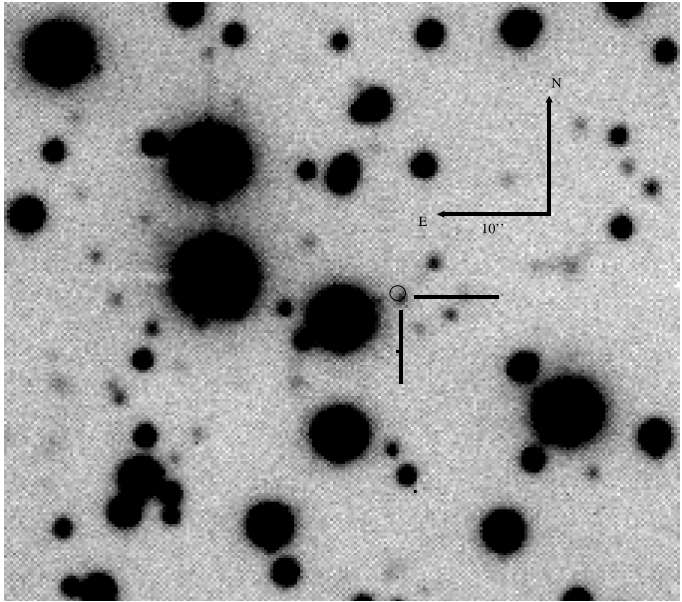, width=9.0cm, height=8.0cm, angle=0}
\vskip 0.0truecm
      \caption{$R-$band finding chart for IGR J00291+5934. The source is indicated by solid bars. The black circle represents the $Chandra$ error box.
              }
         \label{fig:fc}
   \end{figure}

Our entire dataset consist of about $20-30$ frames for each filter (see in Tab.~\ref{tab:log}).
To increase the signal to noise ratio, we computed an average of all our images for each photometric 
band. In all our $V$, $R$ and $I$ averaged frames we clearly detect an object inside the 0.6 arcsec radius Chandra 
error box (\cite{Pa05}), while in our $J$ and $H-$band averaged frame we only have a marginal detection for this object. 
A finding chart is reported in Fig.~\ref{fig:fc}. The position of the detected source is R.A. = 00:29:03.07, Dec. = +59:34:19.12 (J2000) 
with an uncertainty of $0.4''$. This position is coincident with the one of the optical counterpart of IGR J00291+5934 detected by 
\cite{Fo04} during the December 2004 outburst (with an error of $0.5''$) and by Torres et al. (2007) during both outburst and quiescence 
(with an error of $0.05''$) and consistent with the radio position reported by \cite{Ru04}, 
with an uncertainty of less than $0.1''$. Results of PSF-photometry for this source in our optical and NIR averaged frames 
are reported in Tab.~\ref{tab:phot}. We note that our $R-$band magnitude value, as expected, is consistent with the one reported by Torres 
et al. (2007) measured during quiescence, in late 2005 (about two months after our observations). On the other hand, as reported in 
Tab.~\ref{tab:phot}, we did not detect any NIR counterpart of IGR J00291+5934 in $K-$band down to a limiting value of 19.3 mag 
(3$\sigma$ confidence level). We note that Torres et al. (2007) reported a detection at $K = 19.0 \pm 0.1$ for the 2005 Jan 24 observation, about fifty days after 
the discovery of the source in outburst and about twenty days after its return to quiescence (\cite{Jo05}). A possible explanation of this 
discrepancy is that the $K-$band observations of Torres et al. (2007) took place when the source was in the tail of the outburst, and quiescence had not been reached yet.

\begin{table}
\caption{Results of photometry of IGR J00291+5934, all the values are uncorrected for reddening. In column five are reported the dereddening parameters used 
to correct our optical/NIR photometry of computed assuming $E(B-V)=0.74 \pm 0.07$ mag (see Sec. 5 for details).}
\begin{tabular}{ccccc}
\hline
Filter &${\lambda}_c$&  Total Exposure &  Magnitude            &  A$_{\lambda}$  \\ 
       &   (\AA)     &      (s)        &		       &     (mag)       \\ 
\hline
$V$    &  5270       &  5980           & $24.0\pm 0.1$	       & 2.39            \\
$R$    &  6440       &  6240           & $23.2\pm 0.1$	       & 1.79            \\
$I$    &  7980       &  5980           & $22.4\pm 0.2$	       & 1.28            \\
$J$    &  12700      &  1800           & $21.4\pm 0.3$	       & 0.58            \\
$H$    &  16300      &  1500           & $20.4\pm 0.2$	       & 0.39            \\
$K$    &  22000      &  1500           & $\geq 19.3 (3\sigma)$ & 0.25            \\
\hline
\end{tabular}
\label{tab:phot}
\end{table}

\begin{figure}
\epsfig{file=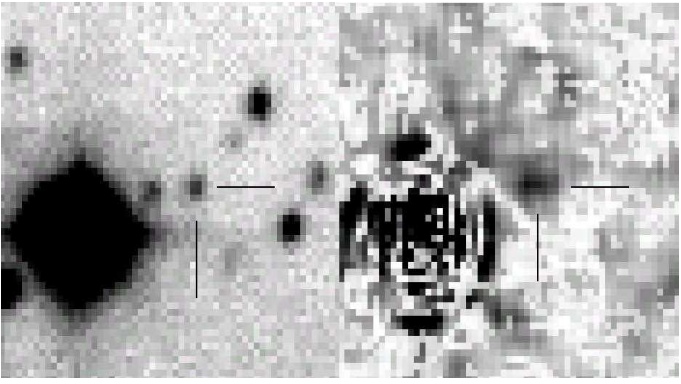, width=9.0cm, height=5.0cm, angle=0}
\vskip 0.0truecm
      \caption{$I-$band Image subtraction for IGR J00291+5934. Left image shows the field of IGR J00291+5934, with its optical counterpart 
      marked. Right image shows the result of the subtraction. A clear residual is present at the position of our source. 
              }
         \label{fig:isis1}
   \end{figure}

\begin{figure}
\epsfig{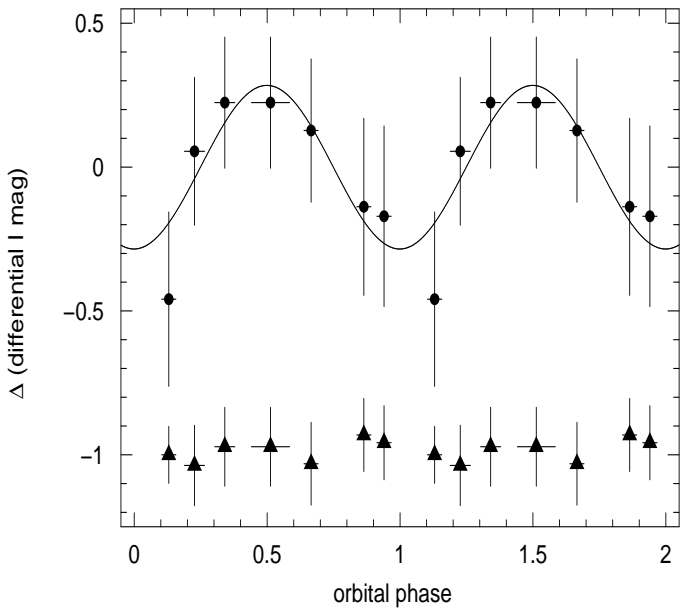}
\vskip 0.0truecm
      \caption{Light curve of IGR J00291+5934 (dots) obtained with image subtraction technique (see Sec. 4 for details) in the phase range 0.10$-$0.97. Two orbital phases 
      are shown for clarity. The best sine$-$wave fit is also shown. The light curve for a field star of comparable brightness is plotted offset below (triangles).
              }
         \label{fig:isis2}
   \end{figure}

\begin{figure}
\epsfig{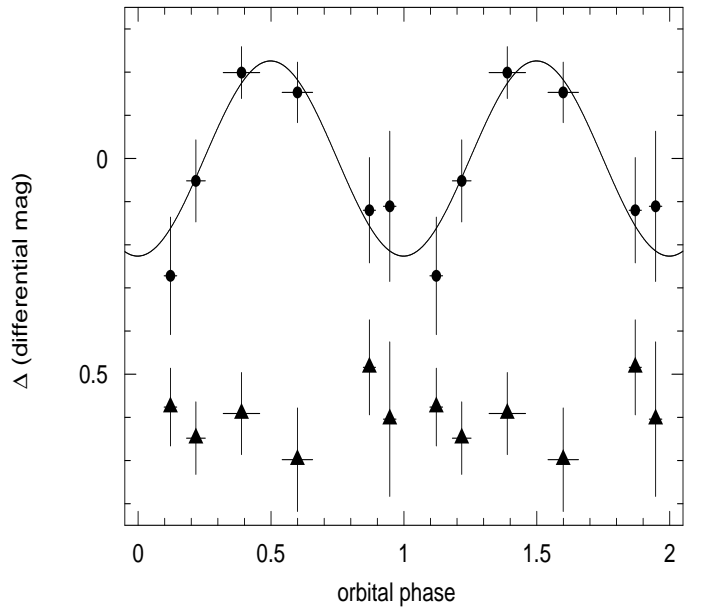}
\vskip 0.0truecm
      \caption{Light curve of IGR J00291+5934 (dots) obtained with PSF$-$photometry of a set of $R$ and $I$ band frames coadded into six phase bins 
      in the phase range 0.10$-$0.97. Two orbital phases are shown for clarity, the best sine$-$wave fit is also shown. The light curve for a 
      field star of comparable brightness is plotted offset below (triangles).
              }
         \label{fig:lc}
   \end{figure}

Once we identified the candidate, we searched for variability 
given that, as reported in Sec. 3, we cover about $80-90$\% of the orbital period with each optical filter. 
As a first check, we performed image subtraction with the ISIS package (\cite{Al00}; \cite{Al98}) on our $I-$band 
images coadded into seven bins. We chose the $I-$band frames because they were those taken under better seeing conditions.
The ISIS subtraction routine accounts for variation in the stellar PSF. A ``reference frame'' (in our case an average of 
two images taken at orbital phases 0.11 and 0.15) is subtracted to all the available images and photometry is performed on the residual images.
For any variable object in the field the variation in flux, with respect to the reference frame, is given as output. 
Performing photometry on the reference frame, it is possible to calibrate in magnitudes the flux variations.
The result of the image subtraction analysis is that our candidate was variable (Fig.s~\ref{fig:isis1}, \ref{fig:isis2}), with an indication of a sinusoidal 
modulation of semiamplitude $0.28 \pm 0.17$ mag (68\% confidence level) at the 2.46 hr orbital period (Fig.~\ref{fig:isis2}). 
The light curve shows a single minimum at phase $0$, i.e. at superior conjunction (when the neutron star is behind the companion) and 
a maximum at phase $0.5$ (based on the X--ray ephemerides of \cite{Ga05}). However, an F$-$test gives a probability of 94.2\% 
with respect to a constant (period and phase constrained) so this represents only a marginal indication of variability of the source.
In light of this, to improve our light curve, we tried to perform a more accurate phase-resolved photometry of our candidate in 
all our optical bands, where we have a clearer detection than in the NIR filters.
Unfortunately, in none of the single optical frames our target was detected with signal-to-noise ratio high enough to look for variability with 
PSF$-$photometry and, considering the uncertainties in the measure of the magnitudes, we can only put a rough upper limit of about 0.3 mag 
on the amplitude of variations in the individual bands.
So, in order to improve the signal-to-noise ratio, we coadded our $R$ and $I-$band frames into 
six phase bins, with a mean 0.05 phase resolution and with all bins containing the same number of $R$ and $I$ band frames, in
order to avoid color-related effects. We discarded $V-$band images because they were taken under worse seeing conditions.
For each resulting bin we performed PSF-photometry of our source and of a selected sample of bright, non-saturated, isolated stars
assumed to be non-variable. The result of such differential photometry of the coadded $R$ and $I-$band frames confirms 
the variability of the source. Assuming that the observed variability could be related to orbital variations, we obtained a good agreement 
by fitting our data with a sinusoid of period 2.46 hr (an F$-$test gives a probability of 99.8\% with respect to a constant, period and phase constrained). 
The resulting semiamplitude is $0.22 \pm 0.09$ mag (68\% confidence level, see Fig.~\ref{fig:lc}).
This unambiguously identifies this source as the optical counterpart of IGR J00291+5934 and represents the first detection 
at both optical and NIR wavelengths of this source during quiescence. Further phase resolved photometry with larger telescopes 
is necessary to study the variability of the source in different bands.

%

\section{Discussion}

Based on X$-$ray pulses arrival times, Galloway et al. (2005) computed for IGR J00291+5934 a mass function of $2.8 \times 10^{-5}$ $M_{\odot}$. This value can be combined
with a system inclination $\leq 85^{\circ}$, derived from the lack of X$-$ray eclipses or dips, and implies a minimum mass for the secondary star of 0.04 
$M_{\odot}$. Assuming that the companion fills its Roche lobe and considering an isotropic, random distribution of inclinations, the companion star must
have a mass $\leq 0.16$ $M_{\odot}$. Such mass limits imply that the mass donor in IGR J00291+5934 is likely a brown dwarf heated by X$-$ray quiescent 
emission of the compact object (\cite{Ga05}). A more tight constrain on the donor mass has recently been derived by Torres et al. (2007). Those authors measured
the peak-to-peak separation of the H$\alpha$ emission profile in the optical spectrum taken during outburst and use this value to estimate the inclination of
the system. They estimate a range of inclination $22^{\circ} \leq i \leq 32^{\circ}$  which implies a mass for the companion star of $M_c = 0.04 - 0.11$
$(0.09 - 0.13)$ $M_{\odot}$ assuming a neutron star of $M_X = 1.4$ $(2.0)$ $M_{\odot}$. Information about the companion's mass can also be obtained from our color photometry. 
To this end, we computed the unabsorbed $(V-R) = 0.2 \pm 0.1$, $(V-J) = 0.8 \pm 0.3$, $(V-H) = 1.6 \pm 0.2$, $(R-J) = 0.6 \pm 0.3$ and 
$(R-H) = 1.4 \pm 0.2$ colors and compared them with the theoretical mass-color diagrams computed for solar metallicity low-mass stars reported in 
Baraffe et al. (1998). The result is that the secondary's mass should be $M_c \geq 0.8$ $M_{\odot}$ ($2\sigma$ c.l.). Using the X$-$ray mass function 
(\cite{Ga05}), we can derive for these masses the inclination of the system $i \leq 3^{\circ}$. The inferred inclination would thus be very low, and therefore 
highly improbable. A similar measure is reported by Torres et al. (2007) from the unabsorbed $(R-K)$ color of IGR J00291+5934. 


The discrepancies in the estimate of the companion's mass can find an explanation under the hypothesis that the
secondary star is subjected to irradiation from the compact object, as suggested by the optical light curve. 
Since during our observations IGR J00291+5934 was in quiescence, we try to 
investigate the possible causes of its optical/NIR emission. Assuming for the source a distance of 3 kpc (as discussed in Sec. 2), then from an observed 
unabsorbed X$-$ray flux of $7 \times 10^{-14}$ erg s$^{-1}$ cm$^{-2}$ measured during quiescence with $Chandra$ (\cite{To07}), we can derive a quiescent 
X$-$ray luminosity of L$_X \sim 8 \times 10^{31}$ erg s$^{-1}$. A broadband spectrum, from optical to NIR, of IGR J00291+5934 can be obtained from our
multiwavelength photometry.
To compute the spectral energy distribution of the source we first need to correct our magnitudes for 
interstellar absorption computed with the relation $N(H)/E(B-V) = 5.8 \times 10^{21}$ 
cm$^{-2}$ mag$^{-1}$ (\cite{Boh79}) and assuming $N(H) = (4.3 \pm 0.4) \times 10^{21}$ cm$^{-2}$ (\cite{Pa05}). The resulting color excess is 
$E(B-V)=0.74 \pm 0.07$ mag. Using a standard extinction curve from \cite{Fi99} we thus obtained the dereddening parameters for each of our 
optical/NIR filters (Tab.~\ref{tab:phot}). 

\begin{figure}
\epsfig{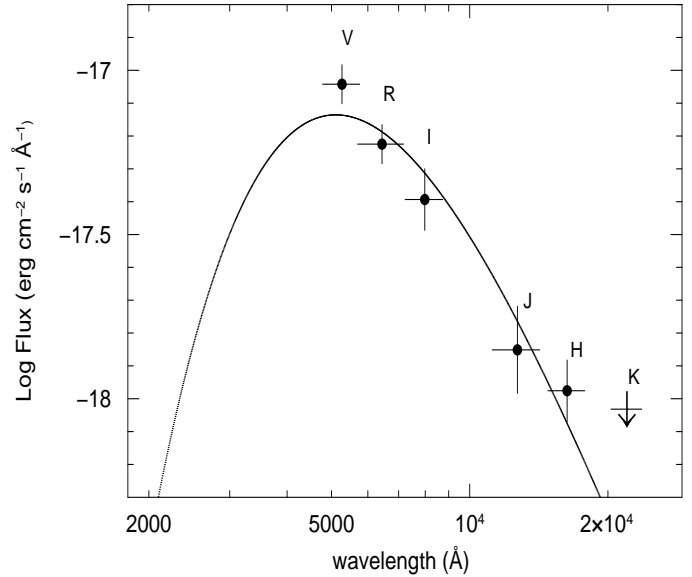}
\vskip 0.0truecm
      \caption{Spectral energy distribution (optical to NIR) of IGR J00291+5934 during quiescence. Data are corrected for interstellar
      absorption. The plotted line represents the contribution of an irradiated star. See Sec. 5 for details.
              }
         \label{fig:sed2}
   \end{figure}

Following Campana et al. (2004) and Burderi et al. (2003) we now attempt to account for the optical to NIR spectral energy distribution (corrected
for interstellar absorption) with the simple model of an irradiated star with a blackbody spectrum (for the details of the modeling see Chakrabarty 1998, 
eqs. [8] and [9]). We fit the data by using the irradiating luminosity ($L_{irr}$) as free parameter and fixing the albedo of the star (${\eta}_* = 0$). We obtain an
acceptable fit to all the data (see Fig.\ref{fig:sed2}), with a reduced ${\chi}^2 = 1.3$ (4 degrees of freedom, null hypothesis probability of 28\%). 
The result of our fit is that the required irradiating luminosity is $4 \times 10^{33} \leq L_{irr} \leq 5 \times 10^{33}$ erg s$^{-1}$. 
Taking this value as a lower limit for the spin-down luminosity of a classical rotating magnetic dipole, we can estimate a neutron star's magnetic field greater than $6 \times 10^7$ Gauss.
The presence of a disc is not required by our fit. In any case, we tried to fit our data with the model of an irradiated star plus a disc by using the 
irradiating luminosity and the internal disc radius ($R_{in}$) as free parameters. We obtained an acceptable fit for similar values of the 
irradiating luminosity ($L_{irr} \sim 3 \times 10^{33}$ erg s$^{-1}$, reduced ${\chi}^2 = 1.8$ with 3 d.o.f., null hypothesis probability of 14\%), 
but the internal disc radius resulting from the fit is not much larger than neutron star ($R_{in} \leq 10^6$ cm). So, we cannot say much about 
the presence of a residual disc structure in the system during quiescence but, according to our results, it is not required to explain the observed 
optical/NIR quiescent emission.

The required irradiating luminosity is about two orders of magnitude larger than the observed quiescent X$-$ray 
luminosity. Such discrepancy is reminiscent of that observed for SAX J1808.4$-$3658 (\cite{Bu03}; \cite{Ca04}), and can be explained with the
presence in the system of a relativistic particle wind from an active pulsar which irradiates the companion star.


%

\section{Conclusions}

We have reported the detection of the optical and NIR counterpart of the accreting millisecond X$-$ray pulsar 
IGR J00291+5934 in quiescence. We provided precise astrometry and the results of multiband ($VRIJHK$) photometry.
Photometry performed on I$-$band images and on a set of coadded $R$ and $I$ band frames, covering about $80-90$\% of the 2.46 hr orbital period, 
shows variability of the source consistent with sinusoidal modulation at the orbital period with a semiamplitude of about 0.2$-$0.3 mag, suggesting the presence 
of an irradiated companion. From the observed $N_H$, we estimate a color excess of $E(B-V)=0.74 \pm 0.07$ mag. The observed optical/NIR quiescent 
luminosity, derived from our multiband photometry, can be modeled with an irradiated companion star providing a good agreement with the observed data. 
The presence of a residual disc component, even if possible, is not necessary for our model.
The required irradiating luminosity is $4 \times 10^{33} \leq L_{irr} \leq 5 \times 10^{33}$ erg s$^{-1}$, much larger than the observed X$-$ray luminosity 
in quiescence. If we assume this value as a lower limit for the spin-down luminosity, then the magnetic field of the neutron star results to be $> 6 \times 10^7$ Gauss. 
As remarked for SAX J1808.4$-$3658 by Burderi et al. (2003) and Campana et al. (2004), the only source of energy available within the system that 
can supply such luminosity is the rotational energy of the neutron star emitted in the form of a relativistic particle wind. However, a direct detection of millisecond 
pulsations in the radio band for IGR J00291+5934 could be difficult for free$-$free absorption effects, due to the mass coming from the companion star and 
swept away by the radiation pressure of the pulsar. A search at high frequencies could be a solution to overcome this
effect (see also Campana et al 1998, Burderi et al. 2003 and Campana et al. 2004).

\begin{acknowledgements}
We thank Vania Lorenzi, Giovanni Tessicini, Noemi Pinilla Alonso and Albar Garcia de Gurtubai Escuder for performing the observations at TNG.
SC and PDA acknowledge the Italian Space Agency for financial support through the project ASI I/R/023/05. PDA acknowledge Daniele Malesani for useful discussion.
\end{acknowledgements}

\end{document}